\documentclass[twocolumn,showpacs,preprintnumbers,amsmath,amssymb,pra]{revtex4}
\usepackage{txfonts}
\usepackage{graphicx}
\usepackage{dcolumn}
\usepackage{bm}
\begin{document}


\title{Controlling directed atomic motion and second-order tunneling of a spin-orbit-coupled atom in optical lattices}
\author{Xiaobing Luo$^{1,2}$}
\altaffiliation{Corresponding author: xiaobingluo2013@aliyun.com}
\author{Zhao-Yun Zeng$^{2}$}
\author{Yu Guo$^{3}$}
\author{Baiyuan Yang$^{2}$}
\author{Jinpeng Xiao$^{2}$}
\author{Lei Li$^{2}$}
\altaffiliation{Corresponding author: fanrenlilei@126.com}
\author{Chao Kong$^{4}$}
\author{Ai-Xi Chen$^{1}$}
\altaffiliation{Corresponding author: aixichen@163.com}
\affiliation{$^{1}$Department of Physics, Zhejiang Sci-Tech University, Hangzhou, 310018, China}
\affiliation{$^{2}$School of Mathematics and Physics, Jinggangshan University, Ji'an 343009, China}
\affiliation{$^{3}$Hunan Provincial Key Laboratory of Flexible Electronic Materials Genome Engineering, School of Physics and Electronic Science, Changsha University of Science and Technology, Changsha 410114, China}
\affiliation{$^{4}$College of Electronic Information and Electrical Engineering, Xiangnan University, Chenzhou 423000, China}
\date{\today}

\begin{abstract}
We theoretically explore the tunneling dynamics for the tight-binding (TB) model of a single spin-orbit-coupled atom trapped in an optical lattice subjected to lattice shaking and to time-periodic Zeeman field. By means of analytical and numerical methods, we demonstrate that the spin-orbit (SO) coupling adds some new results to the tunneling dynamics  in  both multiphoton resonance and far-off-resonance parameter regimes. When the driving frequency is resonant with the static Zeeman field (multi-photon resonances), we obtain an unexpected new
 dynamical localization (DL) phenomenon where the single SO-coupled atom  is restricted to making perfect two-site Rabi oscillation accompanied by spin flipping.
 By using the unconventional DL phenomenon, we are able to generate a ratchetlike effect which enables directed atomic motion towards different directions and accompanies periodic spin-flipping
under the action of  SO coupling.  For the far-off-resonance case, we show that by suppressing the usual inter-site tunneling alone, it is possible to realize a type of spin-conserving second-order tunneling between next-nearest-neighboring sites, which is not accessible in the conventional lattice system without SO coupling. We also show that simultaneous controls of  the usual inter-site tunneling and the SO-coupling-related second-order-tunneling are necessary for quasienergies flatness (collapse) and completely frozen dynamics  to exist. These results may be  relevant to potential applications such as
spin-based
quantum information processing and design of novel spintronics devices.
\end{abstract}
\maketitle
\section{introduction}
In the last two decades, there have been intense and very broad research activities in periodically driven systems\cite{Hanggi,Bukov,Eckardt1,Silveri}.
It has been shown that time-periodic modulation can be used as a flexible knob to generate new states of matter unreachable in static systems. One seminal example is dynamical localization (DL)\cite{Dunlap}, a phenomenon originally proposed by Dunlap and Kenkre in a tight-binding lattice under the influence of a sinusoidal driving field, upon the occurrence of which a localized quantum particle will periodically return to its initial location following the periodic change of the external field. DL has been shown to be connected with quasienergy band
collapse of the underlying time-periodic system\cite{Holthaus}. According to the results obtained by Dunlap and Kenkre, DL happens if the ratio of driving amplitude  to frequency
matches zeros
of the zeroth-order Bessel function. In the high-frequency regime, DL can result in complete suppression of quantum diffusion in
an infinite periodic system, which is more or less similar to the coherent destruction of tunneling (CDT) originally discussed in a double-well potential\cite{Grossmann}. Long studied theoretically, DL has also been
observed experimentally to date in many physical systems, for example, in one-dimensional transverse
Ising chains\cite{Hegde}, periodically curved arrays of optical waveguides\cite{Longhi1,Iyer}, and cold atoms loaded in shaken optical lattices\cite{Lignier}.

 Research on spin-orbit (SO) coupling has also been an active area, which leads to a number of important condensed matter phenomena and applications like spin-Hall effects\cite{Kato}, topological insulators\cite{Bernevig}, spintronics\cite{Sarma}, and quantum computation\cite{Stepanenko}. Recently, artificial SO
coupling
has been realized in experiments with both bosonic and fermionic ultracold atoms\cite{lin471, wang109, cheuk109, zhang109, huang12, wu354}, which provides a completely new platform for exploring the rich
SO-coupled physics. The
combination of spin-orbit coupling with a periodic lattice potential
has become a new hot focus of recent research, which affords the possibility to study exotic magnetic phenomena difficult
to achieve in conventional condensed
matter systems\cite{Cai,Cole,Radic}. Such a spin-orbit-coupled lattice
system has been experimentally realized with a SO-coupled Bose-Einstein condensate (BEC) loaded
into a shallow one-dimensional optical lattice\cite{Hamner}. When combined with an
optical-lattice (OL) potential, SO coupling can give rise to a novel band structure: flat band, which implies a suppression of
dispersive propagation of the wave packet\cite{Kartashov,Zhang1}. The localization of a SO-coupled particle moving in a quasi-periodic OL potential has also been theoretically investigated\cite{Zhou}.

The introduction of time-periodic perturbations to SO-coupled bosonic systems has drawn much interest in recent years.
In this context, there have been several works proposing methods to
tune\cite{Spielman,Wang} or induce\cite{Rahav} SO coupling
based on driven cold-atom systems. For example, the effective SO coupling for continuous BEC systems can be tuned by periodically modulating the Raman frequency\cite{Zhang2} or by
modulating gradient magnetic fields\cite{Ueda},
which have been successfully realized in experiments\cite{Spielman,Wang}. Furthermore, currently a lot of effort is directed towards manipulations of  dynamical properties
of SO-coupled cold-atom systems with versatile periodic driving protocols. For example, it has been shown that periodic driving can be employed for dynamical
suppression of tunneling for the noninteracting SO-coupled BEC system with a double-well potential\cite{Kartashov2}, and for engineering the nonlinear
modes (such as solitons and vortices) of the SO-coupled condensate\cite{Abdullaev1}. Recently, the spin-dependent DL of SO-coupled cold atomic gas in either a driven bipartite lattice\cite{Hai1} or a driven lattice with localized impurity\cite{Luo1} has been reported, but it must be  stressed that in these studies, SO coupling strength needs to be properly selected so as to realize DL. The DL of a single SO-coupled atom held in a periodic potential subjected to
a weak harmonically varying linear force has also been studied\cite{Kartashov2}. The DL reported in Ref.~\cite{Kartashov2} refers to the continuous model,
and subsequently, the contribution of SO coupling on the flat bands and the DL phenomenon has been explored in the discrete settings induced by the presence of deep
optical lattices\cite{Abdullaev2,Abdullaev3}. On the other hand, we also mention a relevant work on a strongly driven fermion system, in which joint effects of tunneling and spin-orbit coupling  can  result in very slow Rabi-like spin-flipping oscillations for a single-electron (a fermion) trapped in a double quantum dot under periodic perturbation by electric field\cite{Khomitsky}.

In this work, we explore the tunneling dynamics for the
tight-binding (TB) model of a single SO-coupled atom trapped in an optical lattice subjected to lattice shaking and to time-periodic
 Zeeman field. We place our emphasis on what new effects the SO coupling will bring to DL phenomenon in such a system. Quasienergy collapse and resulting suppression of tunneling are demonstrated analytically and numerically for both  resonant and far-off-resonant parameter regimes. For the multi-photon resonance case, where the static Zeeman field is an integer multiple of the driving
frequency, we find that under the DL condition, the system reduces to a chain of  disconnected dimers, each of which is described by a symmetric two-state model without energy bias  due to  the fact that the energy offset created by spin flipping is bridged by  multi-photon resonance effect.
We highlight the differences of this DL effect in the SO-coupled lattice systems with respect to the
conventional lattice systems. When the unconventional DL occurs, the single atom makes perfect two-site Rabi oscillation, which suggests
a scheme to produce directed motion and
thus enables the single atom to be steered to move towards specified direction in a controllable way.
In the meantime, the SO coupling leads to periodic spin-flipping in
the directed transport process.
For the far-off-resonant case where the static
Zeeman field is far from any integer multiple of the driving frequency, we show that
 suppression of the usual inter-site tunneling by means of lattice shaking alone is not enough
to induce completely frozen dynamics (CDT), and proper modulation of Zeeman field  is also necessary for suppression of the second-order tunneling
induced by the presence of SO coupling. The capability to suppress the usual inter-site tunneling alone makes it possible to
observe the second-order
tunneling  without
spin flipping between next-nearest-neighboring sites, which is not accessible in the conventional tight-binding lattice in the absence of SO coupling and thus may
provide a new avenue for directly transporting a spin to its non-nearest-neighboring sites without spin flipping.

\section{Model equation}
 We consider the system of a single ultracold boson with two pseudospin
states ($\uparrow$ and $\downarrow$) hopping on
one-dimensional (1D) optical lattices with synthetic SO coupling.  The effective SO coupling can be implemented
experimentally using two counter-propagating Raman lasers which generate a momentum-sensitive
coupling between the two internal hyperfine states of the same
atom. The SO-coupled
 Hamiltonian for single-particle motion along $x$ direction without external potential can be written
in a gauge-field
form\cite{lin471}:
  $\hat{h}=\frac{(\hat{p}_x-\hat{A})^2}{2M}+\frac{\Omega}{2}\hat{\sigma}_z,$
in which $\hat{p}_x$ is the atomic momentum operator, $M$ is the
atomic mass, $\Omega$ behaves as a Zeeman
field, $\hat{\sigma}_{x,y,z}$ are the usual $2\times2$ Pauli matrices, and the effective SO coupling is embodied
in a synthetic non-abelian vector potential $\hat{A}=-k_R\hat{\sigma}_y$  with the wave number of the Raman laser $k_R$ characterizing
the strength of SO coupling. Such a single-particle Hamiltonian can be used to simulate the corresponding condensed-matter system such as the motion of an electron subject to both the Zeeman field and equal Rashba-Dresselhaus SO coupling.

In the presence of sufficiently strong lattice potentials,
this system can be described by the tight-binding Hamiltonian\cite{Cole,Radic,Zhou,Hai1}
\begin{equation}\label{Hamil}
  \hat{H}=\sum_n\Big[-(\hat{c}_n^{\dag}\hat{T}\hat{c}_{n+1}+h.c.)+f(t)n\hat{c}_n^{\dag}\hat{c}_n
  +\frac{\Omega(t)}{2}\hat{c}_n^{\dag}\hat{\sigma}_z\hat{c}_n\Big],
\end{equation}
where $\hat{c}^{\dagger}_{n}=(\hat{c}^{\dagger}_{n,\uparrow},\hat{c}^{\dagger}_{n,\downarrow}), \hat{c}_{n}=(\hat{c}_{n,\uparrow},\hat{c}_{n,\downarrow})^{T}$ (superscript $T$ stands for the transpose),
$\hat{c}^{\dagger}_{n,\sigma} (\hat{c}_{n,\sigma} )$ describes the creation (annihilation) of a  pseudo-spin $\sigma=\uparrow,\downarrow$
 boson at site $n$ ($n=0,\pm 1,\pm 2,...$), and $h.c.$ means the Hermitian conjugate of previous term. When the
vector potential $\hat{A}$ is included, a quantum particle in the 1D periodic potential
(along $x$ axis) picks up a phase $\varphi_n=-\int_{x_n}^{x_{n+1}}\hat{A}dx=-\hat{A}(x_{n+1}-x_n)=k_R(x_{n+1}-x_n)\hat{\sigma}_y$ upon tunneling from site $n$ to $n+1$.
Using the phase $\varphi_n$ to represent the effect of $\hat{A}$, the hopping matrix  changes to $\hat{T}\equiv v\exp(-i\varphi_n)=v(\cos\alpha-i\hat{\sigma}_{y}\sin\alpha)$ through the Peierls substitution, where $v$ denotes the hopping amplitude in the absence of SO coupling and the dimensionless parameter $\alpha=k_R(x_{n+1}-x_n)=\pi k_R/k_L$ (here, $k_L$ is the lattice wave
number) characterizes the
 effective SO coupling strength. The time-periodic potential which varies linearly with site number $n$ models the optical lattice
shaking with the driving field $f(t)$, which is assumed to be sinusoidal with
frequency $\omega$ and amplitude $f_0$, i.e., $f(t)=f_0\cos(\omega t)$.  We also consider
the combined modulations, and assume that the Zeeman field is periodically varying in
time as $\Omega=\Omega(t)=\Omega_0+\Omega_1\cos(\omega t)$, where $\Omega_0$ is the nonzero static part of the field and $\Omega_1$ is the oscillating amplitude
modulated with
the same frequency $\omega$ of the lattice shaking.

Throughout this paper, $\hbar=1$ is adopted and the dimensionless parameters $\nu, \Omega_0, \Omega_1, \omega, f_0$ are measured in units of the reference frequency $\omega_0=0.01 E_r/\hbar$, with $E_r=\hbar^2k_{R}^2/(2M)$ being the single-photon recoil energy, and time $t$ is normalized in units of $\omega_0^{-1}$.
In the experiments\cite{lin471,Lignier,Eckardt}, the Zeeman field $\Omega$ is set as $-400 \omega_0\sim 400 \omega_0$, the
wavelength of the Raman laser is $\lambda_R=804.1$ nm which gives the recoil frequency $E_r/\hbar=\hbar k_{R}^2/(2M)=22.5$ kHz with $k_R=2\pi/\lambda_R$, the usual inter-site hopping coefficient $\nu$ is on order of several tens of Hz,
and the driving
frequency $\omega$ adjustable between $0\sim 20$ kHz. The rescaled SO coupling parameter $\alpha$ depends on the
directions and wavelengths of the applied Raman lasers.
Thus, the experimentally achievable system parameters can be tuned in a wide range as follows: $\Omega_0, \Omega_1, f_0 \sim \omega \in [0, 100](\omega_0)$, $\nu \sim \omega_0$\cite{lin471,Lignier,Eckardt,Xu,Luo2}.

Employing the Wannier state basis $\{|n,\sigma\rangle\}$, where $|n,\sigma\rangle=\hat{c}^{\dagger}_{n\sigma}|0\rangle$\ represents the state of a spin-$\sigma$  particle occupying a lattice site $n$, we can expand the quantum state of the SO-coupled system as
\begin{eqnarray}\label{Statevector}
   |\psi(t)\rangle=\sum_{n,\sigma}c_{n,\sigma}|n,\sigma\rangle,
\end{eqnarray}
where $c_{n,\sigma}(t)$ indicates the time-dependent probability amplitude of the atom being in state $|n,\sigma\rangle$, and the corresponding
probabilities read $P_{n ,\sigma}=|c_{n,\sigma}(t)|^2$, conserving the normalization condition $\sum_{n,\sigma}P_{n ,\sigma}=1$.
Inserting Eqs.~(\ref{Hamil}) and (\ref{Statevector}) into Schr$\ddot{\textrm{o}}$dinger equation $i\partial_{t}|\psi(t)\rangle=\hat{H}|\psi(t)\rangle$, we obtain the following coupled equations
\begin{align}\label{Dnls1}
  i\frac{dc_{n,\uparrow}}{dt}= &-\nu\Big[\cos\alpha(c_{n+1,\uparrow}+c_{n-1,\uparrow})+\sin\alpha(-c_{n+1,\downarrow}+c_{n-1,\downarrow})\Big]
  \nonumber\\&+nf(t)c_{n,\uparrow}+\frac{\Omega(t)}{2}c_{n,\uparrow},\nonumber\\
  i\frac{dc_{n,\downarrow}}{dt}= &-\nu\Big[\cos\alpha(c_{n+1,\downarrow}+c_{n-1,\downarrow})+\sin\alpha(c_{n+1,\uparrow}-c_{n-1,\uparrow})\Big]
  \nonumber\\&+nf(t)c_{n,\downarrow}-\frac{\Omega(t)}{2}c_{n,\downarrow}.
\end{align}
In Eq.~\eqref{Dnls1}, the terms proportional
to $\cos\alpha$ describe the usual spin-conserving
hopping of boson,
while those proportional to $\sin\alpha$ describe the spin-flipping hopping
 arising from a two-photon
Raman process.

\section{Multi-photon Resonance and two-site spin-flipping tunneling }
Multi-photon resonance is one of notable effects of the periodically driven
system. The appealing concept originated in the prototype system with a quantum particle confined in a driven Wannier-Stark lattice\cite{Korsch}, and later also found generalization and application in many-body bosonic system where multi-photon resonance is possible when the energy of $n$ photons bridges the energy gap created by particle interactions rather than static bias\cite{Creffield}. Here, for the SO-coupled tight-binding model, the multi-photon regime is also investigated, where the static Zeeman field is an integer multiple of the
frequency of the driving field, namely, $\Omega_0=m\omega,(m=1,2,3,...)$.
For the high driving frequency $\omega\gg \nu$ (that is, the driving frequency is still much larger than other natural energy scales characterized by $\nu$), it is useful to derive
the effective time-averaged Hamiltonian by implementing a time averaging method  which has been routinely employed for understanding
the celebrated photon resonance effect. In this case, a static effective Hamiltonian can be obtained by time-averaging of the periodic driving effects, i.e.,
\begin{equation}\label{effHamil}
  \hat{H}_{\rm{eff}}=\frac{\omega}{2\pi}\int_0^{\frac{2\pi}{\omega}}dtS^{-1}[-(\hat{c}_n^{\dag}\hat{T}\hat{c}_{n+1}+h.c.)]S,
\end{equation}
where $S=e^{-iA(t)\sum_ll\hat{c}_l^{\dag}\hat{c}_l-iB(t)\sum_l\hat{c}_l^{\dag}\hat{\sigma}_z\hat{c}_l}$,
 $A(t)=\int_0^tf_0\cos\omega tdt=\frac{f_0}{\omega}\sin\omega t$,
and $B(t)=\int_0^t\frac{\Omega(t)}{2} dt=\frac{\Omega_0}{2}+\frac{\Omega_1}{2\omega}\sin\omega t$.
Performing the integral in Eq.~\eqref{effHamil}, we obtain the effective
Hamiltonian
\begin{align}\label{effHamil2}
  \hat{H}_{\rm{eff}}=&-\nu J_0(\chi)\cos\alpha\sum_n(\hat{c}_n^{\dag}\hat{c}_{n+1}+h.c.)\nonumber\\&-\nu\sin\alpha\sum_n\Big[\hat{c}_n^{\dag}(-J_{-m}^-\hat{\sigma}_++J_{-m}^+\hat{\sigma}_-)\hat{c}_{n+1}+h.c.\Big],
\end{align}
where $\hat{\sigma}_{\pm}=\hat{\sigma}_x\pm i\hat{\sigma}_y$,
and
\begin{equation}\label{Bessel}
 J_{-m}^{\pm}\equiv J_{-m}^{\pm}(\eta)=J_{-m}(\eta\pm\chi),
\end{equation}
with
\begin{equation*}
  \chi=\frac{f_0}{\omega},~\eta=\frac{\Omega_1}{\omega},
  \end{equation*}
and with  $J_\gamma(z)$
being $\gamma$-order Bessel function of variable $z$.

Performing the Fourier transformation,
\begin{equation}
\hat{c}_{n,\sigma}=\frac{1}{\sqrt{N}}\sum_ke^{ikn}\hat{a}_{k,\sigma},
\end{equation}
where $k\in[-\pi,\pi]$ is the quasi-momentum varying across the
 first Brillouin zone and $N$ is the total number of lattice sites, we
 transfer the effective Hamiltonian from
the lattice space into the momentum space,
\begin{align}\label{effHamil3}
  \hat{H}_{\rm{eff}}(k)=&\sum_k\hat{a}_k^{\dag}\hat{h}_{\rm{eff}}(k)\hat{a}_k,
\end{align}
with $\hat{a}_k=(\hat{a}_{k,\uparrow},\hat{a}_{k,\downarrow})^T$, and
\begin{align}\label{h_eff}
\hat{h}_{\rm{eff}}(k)=&-2\nu J_0(\chi)\cos\alpha\cos k\hat{I}+\nu\sin\alpha(J_{-m}^--J_{-m}^+)\cos k\hat{\sigma}_x\nonumber\\&-
\nu\sin\alpha(J_{-m}^-+J_{-m}^+)\sin k\hat{\sigma}_y.
\end{align}
Diagonalization of $\hat{H}_{\rm{eff}}(k)$ gives the energy
dispersion of the two-band model,
\begin{align}\label{energy}
  E(k)=&-2\nu J_0(\chi)\cos\alpha\cos k
\nonumber\\&\pm |\nu\sin\alpha| \sqrt{(J_{-m}^+-J_{-m}^-)^2+4J_{-m}^{+}J_{-m}^-\sin^2 k},
\end{align}
where $\pm$ correspond to the upper and lower branches of the dispersion curves (bands), respectively. In the absence of
SO coupling ($\alpha=0$), we have $E(k)=-2\nu J_0(\chi)\cos k$ and the above
dispersion relation reduces to the one extensively addressed in the conventional driven tight-binding model.

\begin{figure}[htp]
\center
\includegraphics[width=8cm]{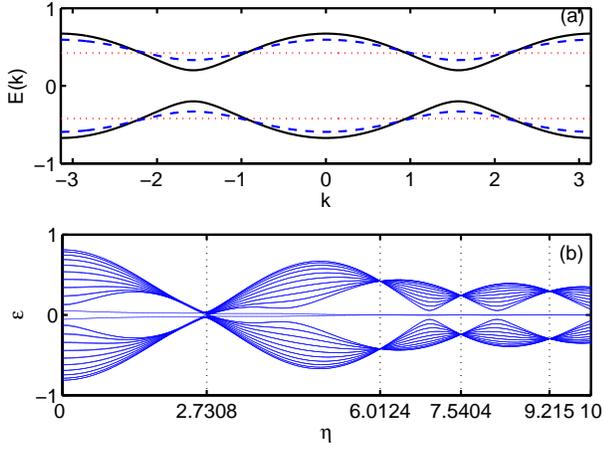}
\caption{(color online) (a) Dispersion curves of the effective time-averaged Hamiltonian (\ref{effHamil2}) [directly obtained from Eq.~(\ref
{energy})] for $\chi=2.408$, and three different modulational
Zeeman field parameters: $\eta=5$ (black solid line), $\eta=5.5$ (blue dashed line), and $\eta=6.0124$ (red dotted line). Other parameters are fixed
as $m=2, \chi=2.408, \nu= 1, \alpha = 0.4\pi$. (b) Numerically-computed quasi-energy spectrum of  the original time-periodic
TB model (\ref{Hamil}) (with a chain of $21$ sites) as a function of the normalized
modulation amplitude of Zeeman field $\eta$. $\Omega_0=2\omega=40$, and the other parameters are the same as in panel (a).} \label{fig1}
\end{figure}

In Fig.~\ref{fig1}(a), we depict the typical dispersion curves obtained from Eq.~(\ref
{energy}) for the resonance case $\Omega_0=2\omega$ (i.e., $m=2$), with fixed $\chi=2.408$ and with three different
values of Zeeman field modulations $\eta$. It is important to note that there exists flattening of
band (red-dotted curve) for certain suitably chosen
values of the modulation parameters $\chi$ and $\eta$, which we will discuss in detail later.

Now we move on to seek for the parameter condition under which the band flatness exists as presented in Fig.~\ref{fig1}(a).
In order to fully suppress the usual inter-site tunneling in the absence of SO coupling ($\alpha=0$), we should first fix  the optical lattice
shaking parameter $\chi=\bar{\xi}$ as a zero of the Bessel function $J_0$,  which means vanishing of the effective spin-conserving hopping between nearest-neighboring sites,
say, $\nu J_0(\chi)\cos\alpha=0$. Under such a circumstance, the lower and upper dispersion bands are
symmetric with respect to $E=0$, and the dependence of $E_\pm(k)$ on the wave vector $k$ is
fully governed by the Zeeman field parameter $\eta$ through the factor $J_{-m}^{+}J_{-m}^-$ in Eq.~(\ref
{energy}).

For the nonzero SO coupling case, it is readily concluded from Eq.~(\ref
{energy}) that under the prerequisite condition $J_0(\chi)=0$, the band flatness can be achieved when one of the equations:
$J_{-m}^{\pm}\equiv J_{-m}(\eta\pm\bar{\xi})=0$
is satisfied, which occurs provided that $\eta$ is taken as
\begin{equation}\label{eta value}
  \eta_l^{\pm}=\beta_l\pm\bar{\xi},
\end{equation}
with $\beta_l(~l=1,2,\dots)$ standing for the $l$th zero of $J_{-m}$ (i.e., $J_{-m}(\beta_l)=0$).

Thus, flat bands
in $k$-space are achieved with the energy
dispersion given by the simple form
\begin{equation}\label{energyd}
  E_\pm=\pm|\nu\sin\alpha||J_{-m}^{\pm}(\eta_l^{\pm})|.
\end{equation}
From Eq.~\eqref{energyd}, it is clear that in the presence of SO coupling ($\sin\alpha\neq 0$),
there is always an energy gap between the two flat bands, while in the absence of SO coupling ($\sin\alpha=0$),
the energy gap closes as $E_\pm=0$. This marks that the SO coupling parameter has
 nontrivial contribution to the characteristic of flat bands.

As an illustration with $\Omega_0=2\omega$ (i.e., $m=2$), we take $\chi=\bar{\xi}=2.408$ (the first zero of $J_0$),
from Eq.~(\ref{eta value}) it follows that the band flatness occurs for the sequence of $\eta$ values
listed in increasing
order as $\{\eta_1^-=\beta_1-\bar{\xi},\eta_2^-=\beta_2-\bar{\xi},\eta_1^+=\beta_1+\bar{\xi},\eta_3^-=\beta_3-\bar{\xi},...\}$ with
$\beta_l(~l=1,2,\dots)$ satisfying $J_{-2}(\beta_l)=0$.
The flatband physics can be understood by exploring the effective velocity of the atomic wave packet, which is defined by the variance of
dispersion with respect to the wave vector, $v_{\rm{eff}}=dE(k)/dk$. Such an extreme
band flattening, $v_{\rm{eff}}=dE(k)/dk=0$, implies that the
dispersive propagation is fully suppressed
and DL occurs.

So far our study is limited to the analytic arguments starting from the time-averaged Hamiltonian (\ref{effHamil2}).
To validate the results obtained with the effective time-averaging approach, we
 plot the numerically-computed quasienergy spectrum of the original time-periodic
model (\ref{Hamil}) versus the dimensionless parameter $\eta$
 in a finite lattice comprising $N=21$
sites for $\Omega_0=2\omega=40, \chi=2.408, \nu=1, \alpha = 0.4\pi$.
 Throughout this paper, numerical simulations on the tight-binding model (\ref{Hamil}) are performed with open boundary conditions. As illustrated in Fig.~\ref{fig1} (b), the quasienergies make up two
separated bands and exhibit a series of band
collapses at which the quasienergies shrink into two distinct degenerate points at the values exactly predicted by Eq.~(\ref{energyd}).
These collapses occur at $\eta=2.7308,6.0124,7.5404,9.2150,...$, which is exactly in correspondence with the sequence of $\eta$ values (hereafter we call these $\eta$ values as DL points)
given by Eq.~(\ref{eta value}). Note that the degenerate
quasienergies at these DL points correspond to the energies of the time-averaged system for different values of $k$ distributed over the
whole Brillouin zone $k\in[-\pi,\pi]$. As compared with the energy curves plotted in $k$-space,
the quasienergy spectrum as illustrated in Fig.~\ref{fig1} (b), due to finite-size effects, shows a pair of additional quasienergies oscillating weakly about
zero, which split off from the two separated bands and physically correspond to
two edge states.

Let us now investigate what unconventional features
SO coupling brings to the dynamics at
the DL points. Two distinct cases for this problem
are listed as follows.
\subsection{\emph{Case I}}
When $J_0(\chi)=0$ and $J_{-m}^+=0$, from Eq.~\eqref{effHamil2}, we have the effective Hamiltonian,
\begin{align}\label{effHamil4}
  \hat{H}_{\rm{eff}}=&
  \nu J_{-m}^-\sin\alpha\sum_n(\hat{c}_n^{\dag}\hat{\sigma}_+\hat{c}_{n+1}+h.c.)
  \nonumber\\=&\nu J_{-m}^-\sin\alpha\sum_n(\hat{c}_{n,\uparrow}^{\dag}\hat{c}_{n+1,\downarrow}+h.c.).
\end{align}
In the Hilbert space  expanded by the Wannier state basis $\{|n,\sigma\rangle\}$, i.e., $\{\cdots,|0,\downarrow\rangle,|0,\uparrow\rangle,|1,\downarrow\rangle,|1,\uparrow\rangle,\cdots\}$,
$\hat{H}_{\rm{eff}}$ becomes a block diagonal matrix
\begin{equation}\label{H mat}
  \hat{H}_{\rm{eff}}=
\left(
  \begin{array}{cccc}
    H_{two-state}^I & 0 & \cdots & 0 \\
    0 & H_{two-state}^I & \cdots & 0 \\
    \vdots & \vdots & \ddots & \vdots \\
    0 & 0 & \cdots & H_{two-state}^I \\
  \end{array}
\right).
\end{equation}
Here $H_{two-state}^I$ represents the $2\times2$ block submatrix in the subspace spanned by only two states
$|n,\uparrow\rangle,~|n+1,\downarrow\rangle$, i.e.,
\begin{equation}\label{H2I}
  H_{two-state}^I=\left(
                   \begin{array}{cc}
                     0 & \nu J_{-m}^-\sin\alpha \\
                     \nu J_{-m}^-\sin\alpha & 0 \\
                   \end{array}
                 \right).
\end{equation}

\subsection{\emph{Case II}}
When $J_0(\chi)=0$ and $J_{-m}^-=0$, the
effective Hamiltonian \eqref{effHamil2} becomes
\begin{align}\label{effHamil5}
  \hat{H}_{\rm{eff}}&=
  -\nu J_{-m}^+\sin\alpha\sum_n(\hat{c}_n^{\dag}\hat{\sigma}_-\hat{c}_{n+1}+h.c.)
 \nonumber\\&= -\nu J_{-m}^+\sin\alpha\sum_n(\hat{c}_{n,\downarrow}^{\dag}\hat{c}_{n+1,\uparrow}+h.c.).
\end{align}
With the Wannier state  basis $\{\cdots,|0,\uparrow\rangle,|0,\downarrow\rangle,|1,\uparrow\rangle,|1,\downarrow\rangle,\cdots\}$, $\hat{H}_{\rm{eff}}$ is also a block diagonal matrix whose diagonal contains blocks of $2\times2$ submatrix
$H_{two-state}^{II}$ in the subspace spanned by two states $|n,\uparrow\rangle,~|n-1,\downarrow\rangle$,
\begin{equation}\label{H2II}
  H_{two-state}^{II}=\left(
                   \begin{array}{cc}
                     0 & -\nu J_{-m}^+\sin\alpha \\
                     -\nu J_{-m}^+\sin\alpha & 0 \\
                   \end{array}
                 \right).
\end{equation}

According to the above analysis, for the SO-coupled lattice system, DL means that the lattice
divides into a chain of disconnected dimers. When DL occurs, a spin $\sigma$ atom tunnels back and forth either between the initially occupied site and its left neighboring site or between the initially occupied site and its right neighboring site, and spin flipping accompanies the  tunneling process.
The tunneling direction depends on which of two equations $J_{-m}^{\pm}=0$ is satisfied.
As mentioned above, the  DL (two-site oscillation) arises for the SO-coupled lattice system, which is intrinsically connected with the two-band flatness and qualitatively distinct from the scenario where the CDT (completely frozen dynamics of the particle) is expected for a particle hopping on a conventional lattice without SO coupling in the  high-frequency regime $\omega>>v$ as considered
in this work.

\begin{figure}[htp]
\center
\includegraphics[width=8cm]{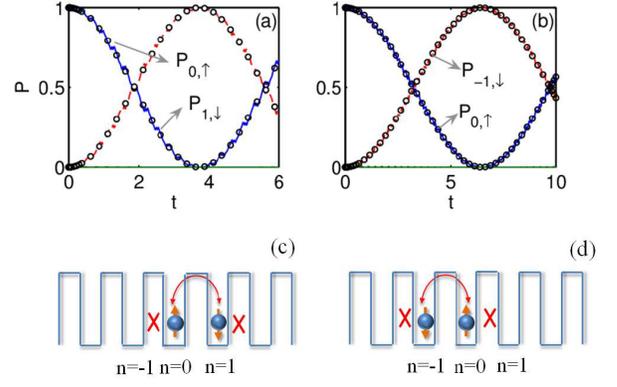}
\caption{(color online) Time evolutions of the occupation probabilities, $P_{n,\sigma} (t )=\langle \psi|\hat{c}^{\dagger}_{n\sigma}\hat{c}_{n\sigma}|\psi\rangle= |c_{n,\sigma} (t )|^2$, in the original SO-coupled TB system
(\ref{Hamil}) (with a chain of $N=21$ sites) for the system initially prepared in state $|0,\uparrow\rangle$. (a) $\chi=\bar{\xi}=2.408, \eta=\beta_2-\bar{\xi}=6.0124$; (b) $\chi=\bar{\xi}=2.408, \eta=\beta_1+\bar{\xi}=7.5404$. The other parameters are set as $\Omega_0=2\omega=40, \nu=1, \alpha = 0.4\pi$. The open circles denote the analytical correspondences obtained from the two-state models (\ref{H2I}) and (\ref{H2II}). Lower panels (c) and (d) are schematic
representations of the tunneling dynamics presented in panels (a) and (b)
respectively. Red crosses indicate suppression of tunneling through
that barrier, and the two-headed arrows indicate the allowed Rabi tunneling with spin flipping.} \label{fig2}
\end{figure}

To illustrate the DL effect more clearly, without loss of generality, we start  the system  with
a single spin-up  atom at a certain site
$n=j$, that is, the initial state given by $|\psi(0)\rangle=|j,\uparrow\rangle$.
At the DL points where $J_0(\chi=\bar{\xi})=0$ and $J_{-m}^{+}\equiv J_{-m}(\eta+\bar{\xi})=0$ (i.e., $\eta=\beta_l-\bar{\xi}$), applying the initial
condition to the two-state model (\ref{H2I}) produces the analytical probabilities,  $P_{j,\uparrow}=\cos^2(\omega_1't), P_{j+1,\downarrow}=\sin^2(\omega_1't)$, and $P_{n,\sigma}=0$  for $n\neq j,j+1$, with
$\omega_1'=|\nu J_{-m}^{-}\sin\alpha|=|\nu J_{-m}(\beta_l-2\bar{\xi})\sin\alpha|$ depending on
the driving field and SO
coupling parameter. Correspondingly, the Rabi period  is given by $T_1=\pi/\omega_1'$. Time evolutions of the atomic probabilities
$P_{n,\sigma} (t ) =\langle \psi|\hat{c}^{\dagger}_{n\sigma}\hat{c}_{n\sigma}|\psi\rangle=|c_{n,\sigma} (t )|^2$ based on the original Hamiltonian \eqref{Hamil} are shown in Fig.~\ref{fig2}(a) for the system  initialized
in the state $|0,\uparrow\rangle$, and with the system parameters $\Omega_0=2\omega=40, \chi=\bar{\xi}=2.408, \eta=\beta_2-\bar{\xi}=6.0124,\nu=1, \alpha = 0.4\pi$. The open circles denote the
analytical  correspondences described by two-state model.
The result means the occurrence
of spin-flipping Rabi oscillation restricted inside a dimer as shown schematically in
Fig.~\ref{fig2}(c), in which only periodic oscillations  are allowed between the state $|j,\uparrow\rangle$ (here, we take $j = 0$ as an example) and the state $|j+1,\downarrow\rangle$.
Similarly, when the other DL requirement $J_0(\chi)=0$ and $J_{-m}^{-}\equiv J_{-m}(\eta-\bar{\xi})=0$ (i.e., $\eta=\beta_l+\bar{\xi}$) is satisfied, from Eq.~(\ref{H2II}) with the initial condition $|\psi(0)\rangle=|j,\uparrow\rangle$, we derive the analytical probabilities  $P_{j,\uparrow}=\cos^2(\omega_2't), P_{j-1,\downarrow}=\sin^2(\omega_2't)$, and $P_{n,\sigma}=0$ for $n\neq j,j+1$, with Rabi period $T_2=\pi/\omega_2'$, $\omega_2'=|\nu J_{-m}^{+}\sin\alpha|=|\nu J_{-m}(\beta_l+2\bar{\xi})\sin\alpha|$. The corresponding  time evolutions of the occupation probabilities are plotted in
Fig.~\ref{fig2}(b) for the
same initial condition and system parameters as in Fig.~\ref{fig2}(a) and yet different $\eta=\beta_1+\bar{\xi}=7.5404$, where the numerical results (see the colored curves) from the full TB Hamiltonian \eqref{Hamil} are in good agreement with the analytical ones (open circles).
They describe spin-flipping Rabi oscillation  along another pathway between the sites $n=j$ and $n=j-1$ as shown schematically in
Fig.~\ref{fig2}(d), where only periodic oscillations (see the two-headed arrow) are possible between the states $|j,\uparrow\rangle$ (here $j=0$) and $|j-1,\downarrow\rangle$.
In the bottom schematic diagrams [panels (c) and (d)], the red crosses in the barriers indicate that the tunneling is not
allowed and the two-headed arrows indicate that the Rabi tunneling with spin flipping is allowed.

\begin{figure}[htp]
\center
\includegraphics[width=8cm]{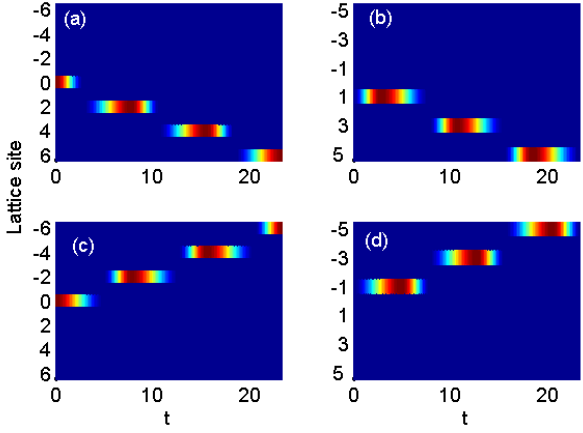}
\caption{(color online) The spatiotemporal evolutions of
$P_{n,\uparrow}(t)=|c_{n,\uparrow}(t)|^2$ (left column) and $P_{n,\downarrow} =|c_{n,\downarrow}(t)|^2$ (right column), obtained from
the original TB system (\ref{Hamil}) (with the size of $13$ sites) for a single spin-up atom initialized at site $n=0$, with the system parameters $\Omega_0=2\omega=40, \chi=\bar{\xi}=2.408, \nu= 1, \alpha = 0.4\pi$. The lighter areas correspond to
the larger values of probability. As shown in the upper row [(a),(b)], the rightward directed motion with periodic spin-flipping is produced by repeating the changes  between $\eta=\beta_2-\bar{\xi}=6.0124$ and $\eta=\beta_1+\bar{\xi}=7.5404$ at the times $t=n(T_1+T_2)/2$ and $t=n(T_1+T_2)/2+T_2/2$ for $n=0,1,2,...$; while as in the lower row [(c),(d)], interchanging the
order of the modulation of the Zeeman field produces directed motion with periodic spin-flipping in
the opposite direction. }  \label{fig3}
\end{figure}

 As mentioned above, by means of a novel
resonance effect between the static Zeeman field and the frequency of the
driving field,  we can realize the DL phenomenon where  the single SO-coupled atom is restricted to making a standard two-site Rabi oscillation (complete population transfer) accompanied with spin flipping.
 In this case,  tunneling is only permitted between two
sites, and the initially localized atom with a given spin can make complete tunneling either to its left
neighbor site or to its right
neighbor site. The tunneling direction is determined by
 which of the two equations  $\eta=\beta_l\pm\bar{\xi}$
is satisfied, where the plus in the plus-minus sign is associated with $J_{-m}^{-}\equiv J_{-m}(\eta-\bar{\xi})=0$ [leading to motion of a spin-up (spin-down) atom initially localized in a given site to the left (right), with the tunneling time $T_2/2=\pi/(2\omega_2')$], and the minus sign corresponds
to $J_{-m}^{+}\equiv J_{-m}(\eta+\bar{\xi})=0$  [leading to motion of a spin-up (spin-down) atom initially localized in a given site to the right (left), with the tunneling time $T_1/2=\pi/(2\omega_1')$].

By make use of the new DL effects illustrated in Fig.~\ref{fig2}, we
can propose a scheme to realize ratchetlike motion of a single atom with spin flipping.
The directed transport (DT) process is illustrated in Fig.~\ref{fig3}, where $\Omega_0=2\omega=40, \chi=\bar{\xi}=2.408, \nu= 1, \alpha = 0.4\pi$. The spin-up particle is loaded
into site $n=0$ and the parameters of the modulational Zeeman field are chosen
as $\eta=\beta_2-\bar{\xi}=6.0124$ satisfying $J_{-m}^{+}\equiv J_{-m}(\eta+\bar{\xi})=0$ initially.
The Rabi
oscillation with spin flipping will occur between the initial
site $n=0$  and its right neighbor site $n=1$, with Rabi period $T_1=\pi/\omega_1'$. At $t=T_1/2$, the particle
arrives at site $n=1$ with spin down. At this time, we immediately adjust the modulational Zeeman field as $\eta=\beta_1+\bar{\xi}=7.5404$ to fit $J_{-m}^{-}\equiv J_{-m}(\eta-\bar{\xi})=0$ such that the particle
enters into a new spin-flipping Rabi oscillation between sites $n=1$ and $n=2$, with the Rabi period $T_2=\pi/\omega_2'$.
When a time interval of $T_2/2$ has elapsed, the
particle has completely tunneled to site $n=2$ and the spin has been flipped back to up.
At the moment,
we tune the driving strength from $\eta=\beta_1+\bar{\xi}=7.5404$ to $\eta=\beta_2-\bar{\xi}=6.0124$; then the single
particle will tunnel from site $n=2$ to site $n=3$ and spin will be again flipped from up to down. If we repeatedly change the $\eta$ values between $\eta=\beta_2-\bar{\xi}=6.0124$ and $\eta=\beta_1+\bar{\xi}=7.5404$ at the times $t=n(T_1+T_2)/2$ and $t=n(T_1+T_2)/2+T_2/2$ for $n=0,1,2,...$, the initial spin-up particle
will propagate solely to the right, with spin flipped periodically in
the transport process. The rightward DT is numerically confirmed from
the original TB model (\ref{Hamil}), as shown in Figs.~\ref{fig3} (a) and (b), where the left (right) panel shows the spatiotemporal evolutions of
the occupation probability of the spin-up (spin-down) component, $P_{n,\uparrow}$ ($P_{n,\downarrow}$), respectively.
Conversely, if the order of the modulation is interchanged,
the initial spin-up particle will propagate solely to the left as shown in
Figs.~\ref{fig3} (c) and (d).  The spatiotemporal evolutions of
$P_{n,\uparrow}(t)=|c_{n,\uparrow}(t)|^2$ [Fig.~\ref{fig3} (c)] and $P_{n,\downarrow} =|c_{n,\downarrow}(t)|^2$ [Fig.~\ref{fig3} (d)] illustrate that the spin is flipped periodically in
the leftward DT process. It is interesting to note that the direction of the
particle's propagation also depends on whether the initial spin is $\uparrow$ or $\downarrow$.
Thus, we can control two reverse spin currents
of the single atom, where a single SO-coupled atom can be
transported toward opposite directions, by adjusting the modulational Zeeman field parameters at some appropriate operation times.
We note that the directed motion was previously proposed by means of a distinct DL mechanism of the driven bipartite lattice\cite{Creffield2}, which nevertheless needs additional choice of the precise values of the SO coupling strength\cite{Hai1}.

The above investigation assumes that the lattice  and  Zeeman
field oscillate in phase with the same
frequency, viz. in Hamiltonian \eqref{Hamil}, $f(t)=f_0\cos(\omega t+\phi)$ and $\Omega(t)=\Omega_0+\Omega_1\cos(\omega t)$, where the
phase difference $\phi$ between the two periodic drivings is set equal to zero. In order to gain more insight into  the nature of the symmetry-breaking
shown in Fig.~\ref{fig2}, we will have a preliminary discussion about  the effect of including a nonzero phase shift $\phi$ on
 the directed motion.  With the same time-averaging procedure as in \eqref{effHamil}, the effective Hamiltonian for general $\phi$ is again given by
 Eq.~\eqref{effHamil2}, but with the substitutions
\begin{align}\label{Sub}
J_{-m}^-\rightarrow H_1,~~~J_{-m}^+\rightarrow H_2,
\end{align}
where we have defined
\begin{align}\label{Sub2}
H_1=\sum_{k=-\infty}^{\infty}J_k(-\chi)J_{-m-k}(\eta)e^{ik\phi}, ~~~H_2=\sum_{k=-\infty}^{\infty}J_k(-\chi)J_{k-m}(\eta)e^{ik\phi}.
\end{align}

By use of the identities, $J_m(z_1+z_2)=\sum_{k=-\infty}^{+\infty}J_k(z_1)J_{m-k}(z_2)$ and $J_k(-z)=(-1)^kJ_k(z)=J_{-k}(z)$, we receive the
following results for two specific cases: (i) When $\phi=0$, then $H_1=J_{-m}(\eta-\chi),~H_2=J_{-m}(\eta+\chi)$;
and (ii) when $\phi=\pi$, then $H_1=J_{-m}(\eta+\chi),~H_2=J_{-m}(\eta-\chi)$. Evidently, in the former case, the results exactly recover the formulas in \eqref{effHamil2} and \eqref{Bessel}, but in the second case, the values of two effective spin-flipping-hopping amplitudes are just interchanged.
For the other values of $\phi$ (namely, $\phi\neq 0,\pi$), $H_1$ and $H_2$ may be complex-valued.
With the same modulational parameters as in Fig.~\ref{fig2} (a), $\chi=\bar{\xi}=2.408, \eta=\beta_2-\bar{\xi}=6.0124$,
we readily derive the  results for the above-mentioned two cases as follows: (i) At $\phi=0$, $H_2$ vanishes, but $H_1$ does not; (ii) at $\phi=\pi$, however, $H_1$  vanishes, but $H_2$  does not. This property implies that the direction of ratchetlike motion can be inverted by only changing the value of $\phi$ from $0$ to $\pi$.

One question naturally raised is: what breaks the left-right symmetry of the lattice, and why applying a
phase shift of $\pi$ between the two drivings inverts the dimerization of the lattice. Here, we will provide a possibly simple explanation for the physical mechanism behind such a phenomenon.
Suppose a spin $\uparrow$ atom tunnels from site $n$ to $n+1$ ($n-1$) via rightward (leftward) spin-flipping hopping,
there will be  a corresponding on-site energies change, $\tilde{B}_1(t)=V_{n,\uparrow}-V_{n+1,\downarrow}=\Omega(t)-f(t)$ [$\tilde{B}_2(t)=V_{n,\uparrow}-V_{n-1,\downarrow}=\Omega(t)+f(t)$] with $V_{n,\sigma}=\pm\Omega(t)/2+nf(t)$ ($+$ for $\sigma=\uparrow$, and $-$ for $\sigma=\downarrow$), respectively, in which the first part of energy change, $\Omega(t)$, comes from spin-flipping, and the second part $f(t)$ from lattice shaking. It behaves as if the rightward (leftward)-hopping atom  moves feeling a combined periodically driving field $\tilde{B}_1(t)=\Omega(t)-f(t)$ [$\tilde{B}_2(t)=\Omega(t)+f(t)$] respectively.
In this case, the driven SO-coupled system behaves
approximately, in a time-averaged sense, like an undriven one, where the tunneling rate from state $|n,\uparrow\rangle$ to $|n+1,\downarrow\rangle$ is
renormalized by a factor of $\frac{\omega}{2\pi}\int_0^{2\pi/\omega}dt\exp[-i\int_0^t dt'\tilde{B}_1(t')]$, and the tunneling rate from state $|n,\uparrow\rangle$ to $|n-1,\downarrow\rangle$ is
renormalized instead by a factor of $\frac{\omega}{2\pi}\int_0^{2\pi/\omega}dt\exp[-i\int_0^t dt'\tilde{B}_2(t')]$.
As we can see, the values of rescaled spin-flipping-tunneling rate are determined by configurations of the combined periodic fields $\tilde{B}_{1,2}(t)$ that the atom feels
[strictly speaking, determined by $\pm\tilde{B}_{1,2}(t)$ for the reason that the Hermitian conjugate hopping term accounting for the reverse hopping process gives rises to the on-site energies change with opposite sign, namely,
the effective tunneling rates occur in complex conjugate pairs]. In principle, owing to $\tilde{B}_{1}(t)\neq\pm \tilde{B}_{2}(t)$, the spin $\uparrow$ atom moving to  its two
neighbors with spin-flipping suffers from different combined periodic fields, and thus breaking of left-right symmetry is to be generically expected.
When the phase difference $\phi=\pi$ is applied between the two drivings, the two combined periodic fields seen by the spin $\uparrow$ atom with spin-flipping tunneling toward two different directions are just swapped as compared with the case of the two in-phase drivings (i.e., $\phi=0$), thereby leading to inversion of  the dimerization of the lattice. Overall, we might intuitively think that the  breaking of left-right symmetry  is a consequence of the combined effect of two drivings. A thorough investigation of the phase-controlled directed motion is worthy to be treated in a future work.

\section{SO-coupling-related second-order tunneling and dynamical localization in far-off-resonant regime}
In the preceding section, we only treat the resonant case. And then a question naturally follows: When the time-periodic modulation is not dealt with in the multiphoton resonance regime, does the phenomenon of suppression of quantum diffusion exist? If it does, what new physical features will be shown? In this section, we are trying to address  these issues. We turn to the off-resonant case where the static Zeeman field $\Omega_0$ is not equal to any integer
multiple of $\omega$, i.e., $\Omega_0\neq m\omega$ for any integer $m$ (absence of
multiphoton resonances). Of particular interest is the far-off-resonant regime with $\Omega_0/\omega=u$ (here, by ``far-off-resonant", we mean that $u$ should take values sufficiently far from any integer) and $\omega\gg \nu$.
In this far-off-resonant regime, the time-averaging treatment becomes invalid, and we will instead perform a multiple-scale asymptotic
analysis \cite{Longhi2,Hai2,Luo3} of the  spin-orbit-coupled bosonic system (\ref{Dnls1}) under two in-phase drivings.  To that end, we start by introducing the normalized  time variable $\tau=\omega t$ and the small parameter $\epsilon=\nu/\omega$. Making the transformation $c_{n,\uparrow}=C_{n,\uparrow}e^{-i\frac{\sin(\omega t)}{\omega}(\frac{\Omega_1}{2}+f_0n)-i\frac{\Omega_0}{2} t},~
  c_{n,\downarrow}=C_{n,\downarrow}e^{-i\frac{\sin(\omega t)}{\omega}(-\frac{\Omega_1}{2}+f_0n)+i\frac{\Omega_0}{2} t}$, and recalling that $\chi=f_0/\omega$ and $\eta=\Omega_1/\omega$, we rewrite Eq.~(\ref{Dnls1}) in terms of the new
amplitudes $C_{n,\sigma}$ as follows
\begin{align}\label{Dnls3}
  i\frac{dC_{n,\uparrow}}{d\tau}=&-\epsilon\cos\alpha\left(e^{-i\chi\sin\tau}C_{n+1,\uparrow}+
  e^{i\chi\sin\tau}C_{n-1,\uparrow}\right)\nonumber\\
  &+\epsilon\sin\alpha
  \left(e^{-i(\chi-\eta)\sin\tau+iu\tau}C_{n+1,\downarrow}-
  e^{i(\chi+\eta)\sin\tau+iu\tau}C_{n-1,\downarrow}\right)\nonumber \\
 i\frac{dC_{n,\downarrow}}{d\tau}=&-\epsilon\cos\alpha\left(e^{-i\chi\sin\tau}C_{n+1,\downarrow}+
  e^{i\chi\sin\tau}C_{n-1,\downarrow}\right)\nonumber\\
  &-\epsilon\sin\alpha
  \left(e^{-i(\chi+\eta)\sin\tau-iu\tau}C_{n+1,\uparrow}-
  e^{i(\chi-\eta)\sin\tau-iu\tau}C_{n-1,\uparrow}\right).
\end{align}

We seek for a solution to Eq.~\eqref{Dnls3} as a power-series expansion in the smallness parameter $\epsilon$:
\begin{align}\label{Series solu}
  C_{n,\uparrow}= &C_{n,\uparrow}^{(0)}+\epsilon C_{n,\uparrow}^{(1)}+\epsilon^2 C_{n,\uparrow}^{(2)}+\cdots, \nonumber \\
  C_{n,\downarrow}= &C_{n,\downarrow}^{(0)}+\epsilon C_{n,\downarrow}^{(1)}+\epsilon^2 C_{n,\downarrow}^{(2)}+\cdots.
\end{align}

At the same time, we introduce multiple  scales for time, $\tau_0=\tau,~\tau_1=\epsilon\tau,~\tau_2=\epsilon^2\tau,\dots$, and then replace the time derivatives by the expansion
\begin{align}\label{Series timederiv}
\frac{d}{d\tau}=\partial_{\tau_0}+\epsilon\partial_{\tau_1}+\epsilon^2\partial_{\tau_2}+\cdots.
\end{align}

Substituting Eqs.~\eqref{Series solu} and \eqref{Series timederiv} into Eq.~\eqref{Dnls3}, we obtain a hierarchy of equations for successive corrections to
$C_{n,\uparrow},~C_{n,\downarrow}$ at different orders in $\epsilon$. At the leading order $\epsilon^0$, we find
\begin{align}\label{0order}
  i\frac{dC_{n,\uparrow}^{(0)}}{d\tau_0}=0,& ~~C_{n,\uparrow}^{(0)}=A_{n,\uparrow}(\tau_1,\tau_2,\cdots),\nonumber \\
  i\frac{dC_{n,\downarrow}^{(0)}}{d\tau_0}=0,& ~~C_{n,\downarrow}^{(0)}=A_{n,\downarrow}(\tau_1,\tau_2,\cdots),
\end{align}
where the amplitudes $A_{n,\uparrow}(\tau_1,\tau_2,\cdots)$ and $A_{n,\downarrow}(\tau_1,\tau_2,\cdots)$ are functions of the slow time variables
$\tau_1,\tau_2,\cdots$, but independent of the fast time variable $\tau_0$. At order $\epsilon^1$ we have
\begin{align}\label{1order}
  i\frac{\partial C_{n,\uparrow}^{(1)}}{\partial\tau_0}=&-i\partial_{\tau_1} A_{n,\uparrow}
  -\cos\alpha\left(e^{-i\chi\sin\tau}A_{n+1,\uparrow}+e^{i\chi\sin\tau}A_{n-1,\uparrow}\right)\nonumber\\
  &+\sin\alpha \left(e^{-i(\chi-\eta)\sin\tau+iu\tau}A_{n+1,\downarrow}-
  e^{i(\chi+\eta)\sin\tau+iu\tau}A_{n-1,\downarrow}\right),\nonumber \\
   i\frac{\partial C_{n,\downarrow}^{(1)}}{\partial\tau_0}=&-i\partial_{\tau_1} A_{n,\downarrow}
  -\cos\alpha\left(e^{-i\chi\sin\tau}A_{n+1,\downarrow}+e^{i\chi\sin\tau}A_{n-1,\downarrow}\right)\nonumber\\
  &-\sin\alpha \left(e^{-i(\chi+\eta)\sin\tau-iu\tau}A_{n+1,\uparrow}-
  e^{i(\chi-\eta)\sin\tau-iu\tau}A_{n-1,\uparrow}\right).
\end{align}
For the convenience of our discussion, we rewrite the above equation \eqref{1order} as
\begin{align}\label{1orders}
i\frac{\partial C_{n,\uparrow}^{(1)}}{\partial\tau_0}=&-i\partial_{\tau_1} A_{n,\uparrow}+K_n^{(1)}(\tau_0),\nonumber \\
   i\frac{\partial C_{n,\downarrow}^{(1)}}{\partial\tau_0}=&-i\partial_{\tau_1} A_{n,\downarrow}+M_n^{(1)}(\tau_0).
\end{align}
To avoid the occurrence of secularly growing terms in the solutions $C_{n,\uparrow}^{(1)}$ and $C_{n,\downarrow}^{(1)}$, the solvability conditions
\begin{equation}\label{solvcondi1}
  i\partial_{\tau_1} A_{n,\uparrow}=\overline{K_n^{(1)}(\tau_0)},~~i\partial_{\tau_1} A_{n,\downarrow}=\overline{M_n^{(1)}(\tau_0)}
\end{equation}
must be satisfied. Throughout our paper, the overline denotes the time average with respect to the fast time variable $\tau_0$. The solvability condition at order $\epsilon^1$ [Eq.~\eqref{solvcondi1}] then gives
\begin{align}\label{solvcondi2}
  i\partial_{\tau_1}A_{n,\uparrow}=&-\cos\alpha J_0(\chi)(A_{n+1,\uparrow}+A_{n-1,\uparrow}),\nonumber\\
  i\partial_{\tau_1} A_{n,\downarrow}=&-\cos\alpha J_0(\chi)(A_{n+1,\downarrow}+A_{n-1,\downarrow}).
\end{align}
According to $C_{n,\uparrow}^{(1)}=-i\int[K_n^{(1)}(\tau_0)-\overline{K_n^{(1)}(\tau_0)}]d\tau_0$ and
$C_{n,\downarrow}^{(1)}=-i\int[M_n^{(1)}(\tau_0)-\overline{M_n^{(1)}(\tau_0)}]d\tau_0$, the amplitudes $C_{n,\uparrow},~C_{n,\downarrow}$
at order $\epsilon$ are given by
\begin{align}\label{1ordersoul}
  C_{n,\uparrow}^{(1)}= & -\cos\alpha F_0(\tau_0)A_{n+1,\uparrow}+\cos\alpha F_0^*(\tau_0)A_{n-1,\uparrow}\nonumber\\&-\sin\alpha F_1(\tau_0)A_{n+1,\downarrow}
            +\sin\alpha F_2(\tau_0)A_{n-1,\downarrow},\nonumber \\
  C_{n,\downarrow}^{(1)}= & -\cos\alpha F_0(\tau_0)A_{n+1,\downarrow}+\cos\alpha F_0^*(\tau_0)A_{n-1,\downarrow}\nonumber\\&-\sin\alpha F_2^*(\tau_0)A_{n+1,\uparrow}
  +\sin\alpha F_1^*(\tau_0)A_{n-1,\uparrow},
\end{align}
where
\begin{align}\label{Frepres}
  F_0(\tau_0)=&\sum_{\substack{p=-\infty \\ p\neq 0}}^{\infty}\frac{J_p(\chi)e^{-ip\tau_0}}{p}, \nonumber \\
  F_1(\tau_0)=&\sum_{p=-\infty}^{\infty}\frac{J_p(\chi-\eta)e^{i(-p+u)\tau_0}}{-p+u}, \nonumber \\
  F_2(\tau_0)=&\sum_{p=-\infty}^{\infty}\frac{J_p(\chi+\eta)e^{i(p+u)\tau_0}}{p+u}.
\end{align}

At the next order $\epsilon^2$, we have
\begin{align}\label{2order}
 i\frac{\partial C_{n,\uparrow}^{(2)}}{\partial\tau_0}=
 &-i\partial_{\tau_2} A_{n,\uparrow}-i\partial_{\tau_1} C_{n,\uparrow}^{(1)}+K_n^{(2)}(\tau_0),\nonumber\\
 i\frac{\partial C_{n,\downarrow}^{(2)}}{\partial\tau_0}=&-i\partial_{\tau_2} A_{n,\downarrow}-i\partial_{\tau_1} C_{n,\downarrow}^{(1)}+M_n^{(2)}(\tau_0),
\end{align}
with \begin{align*}
       K_n^{(2)}(\tau_0)= &-\cos\alpha\left(e^{-i\chi\sin\tau}C_{n+1,\uparrow}^{(1)}+e^{i\chi\sin\tau}C_{n-1,\uparrow}^{(1)}\right)\\
       &+\sin\alpha \left(e^{-i(\chi-\eta)\sin\tau+iu\tau}C_{n+1,\downarrow}^{(1)}
       - e^{i(\chi+\eta)\sin\tau+iu\tau}C_{n-1,\downarrow}^{(1)}\right),\\
       M_n^{(2)}(\tau_0)= &-\cos\alpha\left(e^{-i\chi\sin\tau}C_{n+1,\downarrow}^{(1)}+e^{i\chi\sin\tau}C_{n-1,\downarrow}^{(1)}\right)\\
       &-\sin\alpha \left(e^{-i(\chi+\eta)\sin\tau-iu\tau}C_{n+1,\uparrow}^{(1)}
       - e^{i(\chi-\eta)\sin\tau-iu\tau}C_{n-1,\uparrow}^{(1)}\right).
     \end{align*}
In order to avoid the occurrence of secularly growing terms in the solutions $C_{n,\uparrow}^{(2)}$ and $C_{n,\downarrow}^{(2)}$,
the following solvability conditions must be satisfied:
\begin{align}\label{solvcondi3}
  i\partial_{\tau_2} A_{n,\uparrow}=&\overline{K_n^{(2)}(\tau_0)}\nonumber\\&=-T\sin^2\alpha (A_{n+2,\uparrow}+A_{n-2,\uparrow})+
  \sin^2\alpha(E_1+E_2)A_{n,\uparrow},\nonumber\\
  i\partial_{\tau_2} A_{n,\downarrow}=&\overline{M_n^{(2)}(\tau_0)}\nonumber\\&=T\sin^2\alpha (A_{n+2,\downarrow}+A_{n-2,\downarrow})-
  \sin^2\alpha(E_1+E_2)A_{n,\downarrow},
\end{align}
 where we have set
 \begin{align}\label{Texpr}
   T\equiv & T(\eta,\chi,u)=\sum_{p=-\infty}^{\infty}\frac{J_p(\chi+\eta)J_{-p}(\chi-\eta)}{p+u},\nonumber\\
   E_1=&\sum_{p=-\infty}^{\infty}\frac{J_{p}^2(\chi-\eta)}{-p+u},\nonumber\\
   E_2=&\sum_{p=-\infty}^{\infty}\frac{J_p^2(\chi+\eta)}{p+u}.
 \end{align}
 Thus the evolution of the zeroth-order amplitudes $A_{n,\sigma}$  up to the second-order long time scale is given by
 \begin{equation}\label{AnSecond}
   i\frac{dA_{n,\sigma}}{d\tau}=i\frac{\partial A_{n,\sigma}}{ \partial\tau_0}
   +i\epsilon\frac{\partial A_{n,\sigma}}{\partial\tau_1}+i\epsilon^2\frac{\partial A_{n,\sigma}}{\partial\tau_2}.
 \end{equation}
 Substituting Eqs.~\eqref{0order}, \eqref{solvcondi2} and \eqref{solvcondi3} into Eq.~\eqref{AnSecond}, and returning to the original time
variable $t$, we obtain
\begin{align}\label{Andiff}
    i\frac{dA_{n,\uparrow}}{dt}=&-\nu\cos\alpha J_0(\chi)(A_{n+1,\uparrow}+A_{n-1,\uparrow})
 +\frac{\nu^2\sin^2\alpha}{\omega}(E_1+E_2)A_{n,\uparrow}  \nonumber\\&-\frac{\nu^2\sin^2\alpha}{\omega}T(A_{n+2,\uparrow}+A_{n-2,\uparrow}),\nonumber\\
      i\frac{dA_{n,\downarrow}}{dt}=&-\nu\cos\alpha J_0(\chi)(A_{n+1,\downarrow}+A_{n-1,\downarrow})
   -\frac{\nu^2\sin^2\alpha}{\omega}(E_1+E_2)A_{n,\downarrow}  \nonumber\\&+\frac{\nu^2\sin^2\alpha}{\omega}T(A_{n+2,\downarrow}+A_{n-2,\downarrow}).
\end{align}
 Eq.~\eqref{Andiff} correctly describes the dynamics of the original system \eqref{Dnls1} up to second-order time scale for the far-off-resonant case, where the coefficient $\nu^2\sin^2\alpha T/\omega$ describes
the second-order tunneling rate between two next-nearest-neighboring sites. By neglecting the high-order terms [see Eq.~\eqref{Series solu}], we can approximately view the probability amplitudes $C_{n,\sigma}$
 as the zeroth order $C_{n,\sigma}=C_{n,\sigma}^0=A_{n,\sigma}$.
Thus, $|A_{n,\sigma}|^2$ refers to
the probability of finding a single particle with spin $\sigma$ at
site $n$.  As can be seen from Eq.~\eqref{Andiff}, the dynamics of spin $\uparrow$ is decoupled from
that of spin $\downarrow$, and they are similar to the case of spinless
atom. It is also clearly seen that in
absence of SO coupling ($\sin\alpha=0$), it would be sufficient to fully suppress the
tunneling in the system
by taking $\chi$ as a zero of the Bessel function $J_0$, while in the presence of SO coupling, second-order tunneling remains possible through the SO-coupling-related term $\sin\alpha$ in Eq.~\eqref{Andiff}.

\begin{figure}[htp]
\center
\includegraphics[width=8cm]{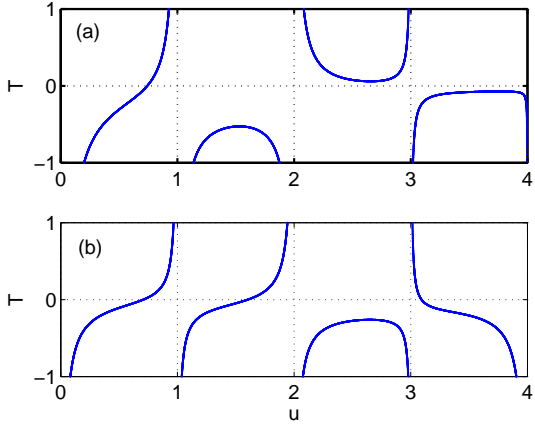}
\caption{(color online) The factor $T$, defined by
Eq.~\eqref{Texpr}, as a function
of $u$, with fixed $\eta=1$, for (a) $\chi=2.4048$, and (b) $\chi=5.5201$.}  \label{fig4}
\end{figure}
Thus, it can be concluded that the frozen dynamics (more properly referred to as CDT) is attained when the two conditions
 \begin{equation}\label{DLcondition_nonre}
 J_0(\chi)=0,~~~T(\eta,\chi,u)=0,
 \end{equation}
are simultaneously satisfied. Notice that the former requirement,
$J_0(\chi)=0$, corresponds to the usual CDT condition in the high-frequency
limit, while the latter represents suppression of second-order process arising from SO coupling.
The numerical calculation shows that $T(\eta,\chi,u)$ can vanish, together with $J_0(\chi)$, for some values
of parameters $\eta,\chi,u$.
This is clearly shown in  Figs.~\ref{fig4} (a) and (b), which describe the factor $T$ of the second-order
tunneling coefficients as a function of the static Zeeman field parameter
$u$ with fixed $\eta=1 $, for $\chi= 2.4048$ and
	 $\chi= 5.5201$ (the lowest two zeros of the $J_0$ Bessel function) respectively.
In these plots, $T(\eta,\chi,u)$ is infinite whenever $u$
takes integer values (and is very large whenever $u$ is close to an
integer).
As a detailed inspection of the  behavior of $T(\eta,\chi,u)$, we find  that in the range $u\in (0,4)$, $T$ has only one zero
point at $u=0.7414$ for $\chi= 2.4048$ (i.e., one has $T=0$ with the parameters $\chi= 2.4048, \eta=1, u=0.7414$), while $T$ has three zero
points at $u=0.7076, 1.6034, 3.0924$ for $\chi= 5.5201$ (at these points, the factor $T$ vanishes). Note that simultaneous vanishing of  $J_0(\chi)$ and $T(\eta,\chi,u)$ means complete suppression of tunneling and hence existence of CDT phenomenon. In the above numerical calculations of $T(\eta,\chi,u)$,  the cut-off value of $p$ is given by $p_{min}=-20$ and $p_{max}=20$, since the sums  converge rapidly and the results remain unchanged if we include the terms of orders higher than the cut-off.

\begin{figure}[htp]
\center
\includegraphics[width=8cm]{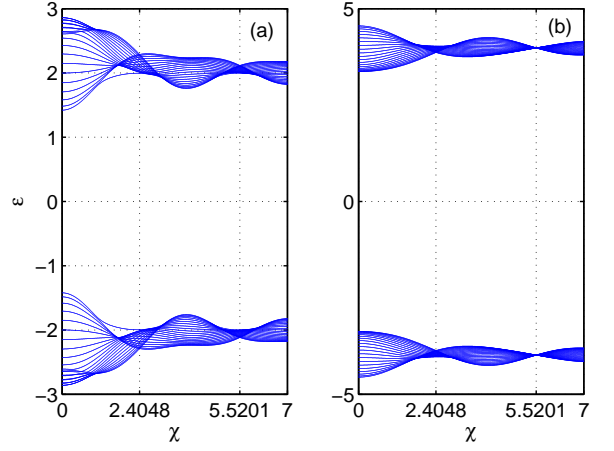}
\caption{(color online) Numerically-computed quasienergies $\varepsilon$ of the time-periodic TB system \eqref{Hamil} comprising $21$ sites as a function of the normalized parameters $\chi$. (a) $\eta=1, u =0.2$; (b) $\eta=1, u =1.6034$. The other parameters are $\omega=20, \nu= 1, \alpha = 0.4\pi$.}  \label{fig5}
\end{figure}

\begin{figure}[htp]
\center
\includegraphics[width=8cm]{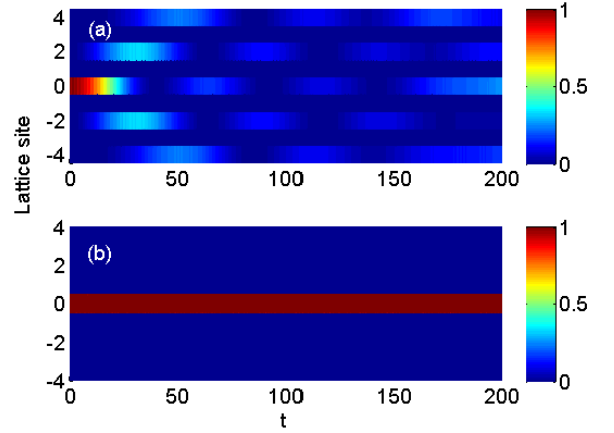}
\caption{(color online) Spatiotemporal evolutions of
the probability $P_{n,\uparrow}$ of  the original TB system \eqref{Hamil} for a spin-up atom initially localized at site $n=0$. (a) $\eta=1, u =0.2, \chi= 5.5201$; (b) $\eta=1, u =1.6034, \chi= 5.5201$. The other parameters are $\omega=20, \nu= 1, \alpha = 0.4\pi$.
The corresponding  occupation
probabilities of spin-down component $P_{n,\downarrow}(t)$ are always negligible during the dynamical evolution and they are not displayed here.}  \label{fig6}
\end{figure}

To confirm the predictions of our asymptotic analysis, we
have numerically calculated the quasienergy spectrum of the time-periodic
TB system \eqref{Hamil} versus the normalized parameter $\chi$
	 in a finite lattice with $21$
sites for $\eta=1$ and for two distinct values of $u = 0.2$ [Fig.~\ref{fig5} (a)]
and $u = 1.6034$ [Fig.~\ref{fig5} (b)]. The other parameters are set as $\omega=20, \nu= 1, \alpha = 0.4\pi$.
Comparing Fig.~\ref{fig5} (a) with Fig.~\ref{fig5} (b),
we can observe a perfect collapse
of quasi-energy bands only at the values of $u = 1.6034, \chi= 5.5201$ which precisely correspond to the second zero of $T(u)$ as shown in Fig.~\ref{fig4} (d). Our results for the spatiotemporal evolutions of the  occupation
probabilities of spin-up component $P_{n,\uparrow}(t)$ are also exemplified in Figs.~\ref{fig6} (a) and (b), by numerical analysis of original TB model \eqref{Hamil}
 with the system parameters as discussed above.
The corresponding  occupation
probabilities of spin-down component $P_{n,\downarrow}(t)$ (not displayed here) are always negligible during the dynamical evolution, due to the fact the usual inter-site hopping is basically suppressed because of $J_0(\chi=5.5201)=0$. As shown in Fig.~\ref{fig6} (a), where $\eta=1, u =0.2, \chi= 5.5201$ [$T\neq 0$ and no collapse
of quasi-energy bands exists, see Fig.~\ref{fig5} (a)], the spin-up atom initialized in site $n=0$ exhibits interesting spatially delocalized propagation.
  In contrast to usual dynamical tunneling, as is clearly shown in Fig.~\ref{fig6} (a),  the initial spin-up atom can be directly transported to its next-nearest-neighboring sites without spin-flipping, which is well consistent with the predictions given by our asymptotic analysis.
The non-spin-flipping tunneling between next-nearest-neighboring sites is a second-order tunneling process, which is a peculiar feature stemming from the SO coupling. Although the validity of asymptotic analysis requires that $u$ should be
 sufficiently far from any integer value, the full-numerical result as presented in Fig.~\ref{fig6} (a) shows that even at $u=0.2$ (being not very far away from the integer $0$), the asymptotic analysis is still a good approximation. This provides some convenience for directly transporting a spin to its
non-nearest-neighboring sites without spin flipping. Conversely, when $\eta=1, u =1.6034, \chi= 5.5201$ (at which the
second-order tunneling factor $T$ vanishes and collapse
of quasienergy bands exists), the spin-up atom remains frozen in its initially occupied site $n=0$ and  CDT occurs as illustrated in Fig.~\ref{fig6} (b).

\begin{figure}[htp]
\center
\includegraphics[width=8cm]{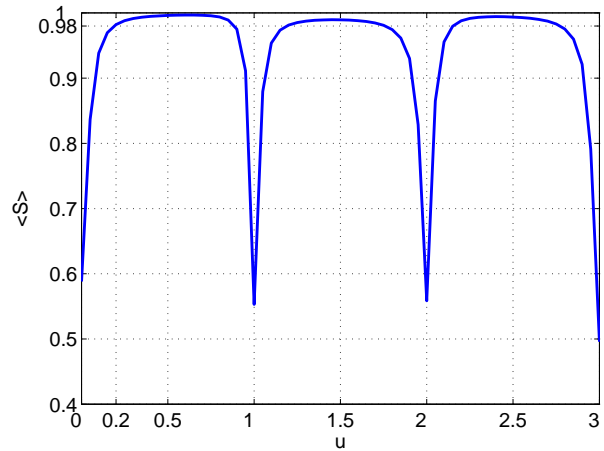}
\caption{(color online) The quantity $\langle S\rangle$  as a function of $u$, with the system initialized in state $|0,\uparrow\rangle$.
Here, $\langle S\rangle$  denotes the
time-averaged total probability of finding the spin $\uparrow$ atom at all the $2n'$ sites with integers $n'=0,\pm 1,...$. The other system parameters are the same as in Fig.~\ref{fig6} (a): $\eta=1,\chi= 5.5201, \omega=20, \nu= 1, \alpha = 0.4\pi$. The time used for averaging is
200 in dimensionless unit.}  \label{fig7}
\end{figure}

Before concluding,  we further identify what parameter ranges of the quantity $u$ can be considered as far-off-resonant regimes where the results based on the multiple-scale asymptotic analysis are applicable. To that end,  we define a related variable $\langle S\rangle=\frac{1}{\Delta t}\int_0^{\Delta t}dt\sum_{n'}|c_{2n',\uparrow}|^2$ with a long-enough averaging time interval $\Delta t$, which is used to measure
the time-averaged total probability of finding the spin $\uparrow$ atom at all the $2n'$ sites with integers $n'=0,\pm 1,...$.
On the condition that the usual spin-conserving tunneling is shut off by taking $J_0(\chi)=0$, and when the system is initialized in state $|0,\uparrow\rangle$,  $ \langle S\rangle= 1$ means that the tunnelings (including the spin-flipping tunneling) between nearest-neighboring sites disappear
which justifies the validity of the multiple-time-scale
asymptotic analysis. As we know, the basic result based on the multiple-time-scale
asymptotic analysis is that the spin-flipping tunneling between nearest-neighboring sites is impossible for any value of system parameters.
Starting the system with a spin $\uparrow$ atom at site 0, from TB model \eqref{Hamil} with size of 13 sites,  we numerically give
$\langle S\rangle$ as the function of $u$ for the other parameters as same as in Fig.~\ref{fig6}. It is
shown in Fig.~\ref{fig7} that the numerical results of $\langle S\rangle$ exhibit different features, as we vary the quantity $u$ to pass through
 the multiphoton resonant points, the near-resonant regions, and the far-off-resonant regions. At each of the resonant points ($\Omega_0=m\omega$ with integers $m$), $\langle S\rangle$ drops to the lowest levels, which indicates the largest nearest-neighbor-tunneling probabilities. As increasing $u$ to be the furthest from resonance (i.e., half-integer values, for example, $m=0.5,1.5,2.5$), $\langle S\rangle$ tends to 1, which indicates disappearance of nearest-neighbor-tunneling as predicted by the multiple-scale asymptotic analysis.
Note that in the near-resonant regions where $\langle S\rangle$ transits from the lowest to near the highest values, the multiple-scale asymptotic analysis is also invalid.
From Fig.~\ref{fig7}, we may estimate that the parameter ranges of $u$ with $\langle S\rangle\geq0.98$ can be regarded as belonging to the far-off-resonant
regime in this paper, since in these regimes the multiple-scale asymptotic analysis works effectively.  As shown in Fig.~\ref{fig7}, the numerical values of $\langle S\rangle$ for the two specific cases exemplified in Fig.~\ref{fig6} can be observed
to be $\langle S\rangle\geq0.98$, indicating that the multiple-scale asymptotic analysis is valid at the values of $u=0.2$ and
$u=1.6034$. The word ``far-off-resonance" is used here for the following two reasons: (i) When $u$ is the furthest from resonance,  the multiple-scale asymptotic analysis works well  without any doubt; and (ii) when $u$ is moderately (but sufficiently) far from resonance, the multiple-scale asymptotic analysis still works, which is  distinguished from the near-resonant regimes.

\section{conclusions}
In summary, we have theoretically studied the tunneling dynamics of tight-binding model for a single SO-coupled atom trapped
in a one-dimensional optical lattice, in the presence of a periodic
time modulation of the Zeeman field and optical lattice shaking. By means of analytical and numerical methods, we have
 evidenced quasienergies flatness (collapse) and suppression of quantum diffusion  for both multi-photon resonance and far-off-resonance cases.
 We have exposed a number
of unconventional features of tunneling dynamics in the presence of SO coupling. For the resonant case, where the static Zeeman field matches an integer multiple of the driving
frequency, we find that as DL occurs the system is divided into a chain of disconnected dimers, each of which is a two-state model that merely couples opposite spins at two neighboring sites. There exists no static bias
inside each two-state model due to the removal of the static Zeeman field by the multi-photon resonance effect.
In
this case the particle is unable to diffuse over the lattice (evidence of DL phenomenon),
and is restricted to performing a perfect two-site Rabi oscillation (the complete population transfer from one site of the dimer to the other) accompanied by spin flipping  under the action of SO coupling. By using the two-site Rabi oscillation of the DL effect that are absent in  conventional lattice system, we are able to generate a  ratchetlike (directed) motion with periodic spin-flipping
of the single SO-coupled atom, in which the direction of motion can be manipulated in a controllable way.  For the far-off-resonant case,
by means of the multiple-time-scale asymptotic analysis, we have revealed that besides the usual non-spin-flipping tunneling between nearest-neighboring sites, there exists a type of second-order tunneling between next-nearest-neighboring sites induced by the presence of SO coupling. It has been shown
that suppression of the usual inter-site tunneling by lattice shaking is not enough
to observe completely frozen dynamics (CDT), and suppression of the second-order tunneling via the Zeeman field modulation is also required.

Finally, we brief discuss the experimental aspects of our results. As implemented in
the seminal work\cite{lin471}, a synthetic
SO coupling can be realized by applying a pair of counterpropagating
Raman beams to couple  two hyperfine
spin states $|\uparrow\rangle=|F = 1,m_F = 0\rangle$ and $|\downarrow\rangle=|F = 1,m_F =-1\rangle$
(with the state $|F=1,m_F=1\rangle$
far detuned from this pseudo-spinor set) of a single ultracold atom. An appropriate optical lattice can
be induced by two additional contrapropagating laser beams at $\lambda_L=1030$ nm wavelength\cite{Eckardt}, with
potential strength of the order $\sim 10E_\textrm{rec}$ [where $E_\textrm{rec}=2\hbar^2\pi^2/(m\lambda_L^2)$],
which guarantees the effectiveness of the tight-binding model. One
can use an acousto-optic modulator to
 precisely
control the lattice beam intensity, and to introduce a
 frequency difference
between the beams that allows us to periodically
shake the lattice\cite{Eckardt}. In addition, the
Zeeman-field intensity can be tuned independently by the
bias magnetic field and the frequency difference between the
two Raman laser beams that couple the two internal atomic states\cite{lin471}.
Alternatively, the SO coupling considered in this work could be realized using the tripod scheme implemented with three laser beams\cite{Edmonds}.
Based on these experiment
setups mentioned above, it should be possible to test our results and in particular the new aspects of DL phenomenon induced by SO coupling.

\acknowledgments
The work was supported by the National Natural Science
Foundation of China (Grants No. 11975110, No. 11764022,
No. 11465009, and No. 11947082), the Scientific and Technological
Research Fund of Jiangxi Provincial Education Department
(No. GJJ180559, No. GJJ180581, No. GJJ180588,
and No. GJJ190577), Open Research Fund Program of the
State Key Laboratory of Low-Dimensional Quantum Physics
(No. KF201903), and the Hunan Provincial Natural Science Foundation of China (Grant No. 2017JJ2272).
Yu Guo was supported by the Scientific Research Fund
of Hunan Provincial Education Department under Grant No. 20A025 and Changsha Municipal Natural Science Foundation under Grant No. kq2007001.

\end{document}